\documentclass[showpacs,aps,twocolumn]{revtex4}
\usepackage{diagbox}
\usepackage{adjustbox}
\usepackage{float}
\usepackage{bm}
\usepackage{amsmath}
\usepackage{graphicx}
\usepackage{subfigure}
\usepackage[usenames,dvipsnames]{color}
\definecolor{darkblue}{RGB}{0,0,196}
\usepackage[colorlinks=true,linkcolor=darkblue,citecolor=darkblue,urlcolor=darkblue]{hyperref}

\usepackage{setspace}
\usepackage{footmisc}
\usepackage[makeroom]{cancel}
\usepackage{comment}
\usepackage{lineno}
\usepackage{bbm}

\def\be{\begin{equation}}
\def\ee{\end{equation}}
\def\ba{\begin{eqnarray}}
\def\ea{\end{eqnarray}}

\usepackage{graphicx}
\usepackage{amsmath,bbm}
\usepackage{amssymb,bm}
\usepackage{yfonts}

\begin{document}
\title{Centrality dependence of Electrical and Hall conductivity at RHIC and LHC energies for a Conformal System}
\author{Bhaswar Chatterjee}
\author{Rutuparna Rath}
\author{Golam Sarwar}
\author{Raghunath Sahoo}
\email{Raghunath.Sahoo@cern.ch}
\affiliation{Department of Physics, Indian Institute of Technology Indore, Simrol, Indore- 453552, INDIA}

\begin{abstract}
\noindent
In this work, we study electrical conductivity and Hall conductivity in the presence of electromagnetic field using Relativistic Boltzmann Transport Equation with Relaxation Time Approximation. We evaluate these transport coefficients for a strongly interacting system consisting of nearly massless particles which is similar to Quark-Gluon Plasma and is likely to be formed in heavy-ion collision experiments.
We explicitly include the effects of magnetic field in the calculation of relaxation time. The values of magnetic field are obtained for all the centrality classes of Au+Au collisions at $\sqrt {s_{\rm NN}} =$ 200 GeV and Pb+Pb collisions at $\sqrt {s_{\rm NN}} =$ 2.76 TeV. We consider the three lightest quark flavors and their corresponding antiparticles in this study. We estimate the temperature dependence of the electrical conductivity and Hall conductivity for different strengths of magnetic field. We observe a significant dependence of temperature on electrical and Hall conductivity in the presence of magnetic field.

\end{abstract}
\date{\today}
\maketitle 
\section{Introduction}
\label{intro}
Relativistic heavy-ion collisions are dedicated to study the quark-gluon plasma (QGP) state of nuclear matter, which is predicted according to Quantum Chromodynamics (QCD)~\cite{Bazavov:2014pvz,Borsanyi:2013bia} and according to the standard model of cosmology, the universe has gone through this phase around few microseconds after the Big bang. An extensive study of the results of heavy-ion collision experiments, by comparing the results with predictions from fluid dynamic models, strongly hints to the formation of QGP in such collisions~\cite{Krzewicki:2011ee,Aamodt:2010cz,Hirano:2010je,Arsene:2004fa,Back:2004je,Adams:2005dq,Shuryak:2004cy,Gyulassy:2004zy,Muller:2006ee}. These studies also conclude that for successful explanations of the experimental results using relativistic hydrodynamic models, the produced QGP requires to have very small shear viscosity to entropy ratio~\cite{Kovtun:2004de,Romatschke:2007mq,Heinz:2013th}. Such comparisons with models, not only confirm the creation of QGP but also gives an opportunity to extract information about various thermodynamic and collective properties of QGP- like equation of state and transport properties of QGP, as they play the role of adjustable parameters in the model. Conditions created in such collisions actually help in extracting these properties, e.g., initial state geometry helps to extract viscosity of QGP, as it causes the final state momentum anisotropy, which gives measure of viscosity to entropy ratio~\cite{Heinz:2013th,Gale:2013da,Teaney:2009qa,Romatschke:2009im,Luzum:2008cw,Song:2007ux,Dusling:2007gi,Molnar:2008xj,Bozek:2009dw,Chaudhuri:2009hj,Schenke:2010rr}. Apart from the initial geometry, some other initial phenomena also significantly affects the way we need to look at it.

One such initial state phenomenon is the creation of magnetic field depending on centrality, atomic number of the nuclei and center-of-mass energy ($\sqrt{s}$)~\cite{Skokov:2009qp}. It has long been proposed that ultra-strong ($\sim 10^{18}$ Gauss) magnetic field can get created in heavy-ion collisions if the number of spectator nucleons is sufficiently high ~\cite{Kharzeev:2007jp,Tuchin:2013ie,Hattori:2016emy}. This is possible in off-central collisions and this can give rise to novel phenomena like chiral magnetic effect as well as new complexities~\cite{Kharzeev:2012ph}. Transport coefficients are extremely important parameters as they determine the evolution of the QGP and generate anisotropic flow velocity which can be measured. Magnetic field, if present, can be another source of added anisotropy while also affecting the phase space of the charged particles and thus the transport coefficients.

One extremely important transport coefficient is the electrical conductivity, $\sigma_{el}$ which significantly affects the dilepton production which is a good probe to investigate QGP~\cite{Ding:2010ga,Moore:2006qn}. The presence of magnetic field can significantly affect this coefficient by affecting the phase spaces and also by generating a force perpendicular to that of the electric field. In such cases, along with regular electrical conduction, Hall conduction may also happen. However, unlike electrical conductivity, the Hall conductivity depends explicitly on the cyclotron frequency of the charged particles. This has serious implications on the outcome as QGP is different from electron-ion plasma where the mobility of oppositely charged particles is different because of their different masses. QGP is a pair plasma where the mobility of oppositely charged particles is the same because of similar masses. So in case of perfectly equal number of oppositely charged particles, the net Hall current vanishes. In most of the high energy heavy-ion collision experiments so far, the baryon chemical potential ($\mu_B$) produced is very small at the RHIC and LHC energies. Also, generation of sufficiently strong magnetic field requires a high center-of-mass energy thus reducing the possibility of high $\mu_B$. But for some values of $\sqrt{s}$ for particular collisions, a non-negligible $\mu_B$ can be created and it would be interesting to observe the Hall conductivity and its interplay with electrical conductivity in such scenarios.

Recently, Hall conductivity has been studied in the context of heavy-ion collision experiments using different methods~\cite{feng2017,Das:2019wjg,Das:2019ppb}. It has been studied in the hadronic phase using hadron resonance gas (HRG) model also~\cite{Das:2019wjg}. In the QGP phase, quasiparticle model has been used in relaxation time approximation to investigate this~\cite{Das:2019ppb}. In this work, we calculate the electrical and hall conductivities by solving Boltzmann transport equation using the relaxation time approximation in presence of magnetic field for a system consisting of nearly massless particles like quarks and gluons while ignoring the QCD interaction between them. In addition to this, we differentiate our work from previous works in the following ways:
\begin{itemize}
\item We explicitly incorporate the effects of magnetic field in the relaxation time itself as well as through the cyclotron frequency ($\omega_c$). This is especially important as a strong magnetic field can significantly affect the scattering processes contributing to the relaxation time. We also take a magnetic field dependent coupling instead of a temperature dependent one which has been used in previous studies \cite{feng2017,Das:2019wjg}.
\item We have considered phenomenologically relevant magnetic fields and baryon chemical potential. In particular, we have worked in the context of two different collision experiments: Pb+Pb collision at $\sqrt {s_{\rm NN}} =2.76$ TeV at LHC and Au+Au collision at $\sqrt {s_{\rm NN}} =200$ GeV at RHIC. We also present a centrality-wise analysis of our results to make them more relevant phenomenologically.
\end{itemize}

But for the sake of simplicity, we have not considered the QCD dynamics which might be relevant at the center-of-mass energies we have considered as the magnetic field produced can be too high depending on the centrality. So, this work can be considered as an introductory 
level study of what happens to a free gas consisting of nearly massless particles like quarks and gluons but without any QCD interaction 
among them.

We have arranged the paper in the following format. In the next section, we present a brief description of the derivation of $\sigma_{el}$ and $\sigma_H$ by solving the Boltzmann transport equation using relaxation time approximation. Then we discuss our results and summarize.

\begin{table*}[htbp]
\begin{center}  
\caption{Magnetic field values in different centrality classes.}
 \begin{adjustbox}{max width=\textwidth}
 \begin{tabular}{||c|c|c|c|c|c|c|c||}
 \hline
\diagbox[width=10em]{eB $\rm (GeV^{2})$}{Centrality ($\%$)} & 0-10 & 10-20 & 20-30 & 30-40 & 40-50 & 50-60 & 60-70\\ 
 \hline
 Au+Au & 0.045 & 0.063 & 0.078 & 0.090 & 0.098 & 0.102 & 0.103 \\ 
 \hline
 Pb+Pb & 0.59 & 0.81 & 0.99 & 1.13 & 1.24 & 1.31 & 1.33 \\ 
 \hline
 \end{tabular}
 \end{adjustbox}
 \label{table:1}
\end{center}
\end{table*}
 
\section{Formulation}
\label{formulation}
\subsection{Electrical Conductivity and Hall Conductivity}
\label{Electrical Conductivity and Hall Conductivity}
We obtain the electrical and Hall conductivities by solving the Boltzmann transport equation using the relaxation time approach in presence of an electromagnetic field as it has been done in Ref.\cite{feng2017,Das:2019wjg}. In presence of an electromagnetic field, the Boltzmann transport equation is given by 
\begin{equation}
 p^{\mu}\partial_{\mu}f(x,p)+q_{f}F^{\mu\nu}p_{\nu}\frac{\partial f(x,p)}{\partial p^{\mu}} = \mathcal{C}[f],
 \label{bte}
\end{equation}
where $q_f$ is the fractional electric charge of the particle, $F^{\mu\nu}$ is the electromagnetic field strength tensor and $\mathcal{C}[f]$ is the collision integral~\cite{Hosoya:1983xm}. In the relaxation time approximation, 

\begin{align}
 \mathcal{C}[f]\simeq -\frac{p^{\mu}u_{\mu}}{\tau}(f-f_0)\equiv -\frac{p^{\mu}u_{\mu}}{\tau}\delta f ,
 \label{rta}
\end{align}
where, $u_{\mu}=(1,\vec{0})$ is the fluid four velocity in the local rest frame and $\tau$ is the relaxation time. For the thermal motion of the particles, here we assume that the magnetic field acts against the equilibration in three dimensions. So, to allow this thermal motion, we consider the particles to have the dispersion relation of a free particle, which is in three dimensional spatial motion. This is because they are not taken to be conditioned by the magnetic field to have effective equilibration in one dimension, disallowing the transverse transports. For the same reason, we have neglected the separation between transverse and longitudinal conductivity. Though at a very high magnetic field longitudinal conductivity will be different from the transverse one. With this assumption, we consider the Boltzmann distribution function satisfying 
\begin{align}
 \frac{\partial f_0}{\partial \vec{p}}=\vec{v}\frac{\partial f_0}{\partial \epsilon}, ~~ \frac{\partial f_0}{\partial \epsilon}=-\beta f_0,
 ~~f_0 = g e^{-\beta(\epsilon\pm\mu)}
 \label{bdf}
\end{align}

 where $\vec{v}$ is the single particle velocity, the single particle energy is $\epsilon(p)=\sqrt{\vec{p}^2+m^2}$, $g$ is the degeneracy, $\mu$ is the chemical potential and $\beta=1/T$, is the 
 inverse of temperature. 
   Now, the Boltzmann transport equation can be written as 
 
  \begin{align}
  \frac{\partial f}{\partial t}+\vec{v}.\frac{\partial f}{\partial\vec{r}}+q_f\bigg[\vec{E}+\vec{v}\times\vec{B}\bigg]
  .\frac{\partial f}{\partial\vec{p}}=-\nu(f-f_0),
  \label{equ4}
 \end{align} 
 where $\nu=1/\tau$. Executing the Fourier transformation of $f(p,x)$, the Boltzmann equation can be rewritten as
 
 \begin{align}
  (-i \omega + i \bf{k}.\bf{v} & +\nu)(f-f_0) = -q_f\bigg[\vec{E}+\vec{v}\times\vec{B}\bigg]\frac{\partial f}{\partial\vec{p}} .
  \label{equ5}
 \end{align}
 Describing $is = i\omega - i \bf{k}.\bf{v} $ , the above equation can be written as
  \begin{align}
  (-is +\nu)(f-f_0) = -q_f\bigg[\vec{E}+\vec{v}\times\vec{B}\bigg]\frac{\partial f}{\partial\vec{p}} .
  \label{equ5A}
 \end{align}

 Without any loss of generality, we take constant electric and magnetic fields to be perpendicular to each other as $\vec{E}=E\hat{x}$ and $\vec{B}=B\hat{z}$. Now, Eq. (\ref{equ5A}) can be written as  
 \begin{align}
 \bigg(-is+\nu-q_fB\bigg(v_x\frac{\partial}{\partial p_y}-v_y\frac{\partial}{\partial p_x}\bigg)\bigg)f(p)\nonumber\\\nonumber
 \\=(-is+\nu) f_0(p) - q_fE\frac{\partial}{\partial p_x}f_0(p). 
 \label{equ6}
\end{align}

In order to solve Eq. \eqref{equ6}, we consider the following ansatz of the distribution function $f(p)$ \cite{feng2017},

\begin{align}
 f(p)=f_0-\frac{1}{-is+\nu}q_f\vec{E}.\frac{\partial f_0(p)}{\partial \vec{p}}-\vec{\Xi}.\frac{\partial f_0(p)}{\partial \vec{p}}.
 \label{equ7}
\end{align}

Using the ansatz given in Eq. \eqref{equ7} and neglecting the higher order terms, Eq. \eqref{equ6} can be simplified as, 
\begin{align}
 \frac{q_fB}{\nu}q_fE\frac{v_y}{p}-q_fB\bigg(v_x\Xi_y-v_y\Xi_x\bigg)\frac{1}{p}\nonumber\\
 +(\nu-is)v. \Xi=0.
 \label{equ9}
\end{align}

Where, the quantity $\Xi$ represents a correction factor that has been used to incorporate the effect of magnetic field in the distribution function which is to be extracted from the solution of Boltzmann solution and the explicit form of $\Xi$ is given in the Eq. \eqref{equ12}. The above equation satisfies for any arbitrary value of velocity which leads to $\Xi_z = 0.$ Further introducing the velocity as $v = v_{x} + iv_{y}$ and $\Xi = \Xi_{x}-i\Xi_{y}$ in Eq. \eqref{equ9}, we obtain
\begin{align}
   Re\bigg[-i\frac{q_{f}E\omega_{c}}{\nu-is}v + (\nu-is-i\omega_{c})v.\Xi]=0,
   \label{sol1}
 \end{align}
where $Re$ means to be the real part and $\omega_c=\frac{q_fB}{\epsilon(p)}$ is the cyclotron frequency. Now the values of both $\Xi_x$ and $\Xi_y$ are obtained as

\begin{align}
 \Xi_x=-q_fE\frac{\omega_c(\omega_{c}+2s)\nu}{(\nu^{2}-s(s+\omega_{c}))^{2}+\nu^{2}(2s+\omega_c)^2},\nonumber\\\nonumber
 \\ ~~~\Xi_y=-q_fE\frac{\omega_c(\nu^{2}-s(s+\omega_{c}))}{(\nu^{2}-s(s+\omega_{c}))^{2}+\nu^{2}(2s+\omega_c)^2}.
 \label{equ12}
\end{align}

In the static limit ($s = (\omega-\vec{k}.\vec{v}) \rightarrow 0$), the ansatz for the distribution function $f(p)$ in Eq. \eqref{equ7} is simplified using Eq. \eqref{equ12} and given by,

\begin{align}
 f(p) & =f_0-q_{f}E\frac{p_x}{p}\frac{\partial f_0}{\partial p}\frac{\nu}{\nu^2+\omega_c^2}
 +q_{f}E\frac{p_y}{p}\frac{\partial f_0}{\partial p}\frac{\omega_c}{\omega_c^2+\nu^2}\nonumber\\
 & =f_0-q_{f}Ev_x\left(\frac{\partial f_0}{\partial \epsilon}\right)\frac{\nu}{\nu^2+\omega_c^2}
 +q_{f}Ev_y\left(\frac{\partial f_0}{\partial\epsilon}\right)\frac{\omega_c^2}{\omega_c^2+\nu^2
 }
 \label{sol_bte}
\end{align}

The electric current can be written in the following form \cite{feng2017}, 

\begin{align}
 j^i=q_{f}\int\frac{d^3p}{(2\pi)^3}v^i\delta f =\sigma^{ij}E_j=\sigma^{el}\delta^{ij}E_j+\sigma^H\epsilon^{ij}E_j,
  \label{current}
\end{align}

For an isotropic system, comparing eq.\ref{sol_bte} and eq.\ref{current}, the electric and the Hall conductivities can respectively be expressed as: 

\begin{align}
 \sigma^{el}=\sum_i
 \frac{q_{fi}^2\tau_i}{3T}\int\frac{d^3p}{(2\pi)^3}\frac{p^2}{\epsilon^2_i} \frac{1}{1+(\omega_{ci}\tau_i)^2}f_{0i},
 \label{sigma_el}
\end{align}

\begin{align}
 \sigma^{H}= \sum_i\frac{q_{fi}^2\tau_i}{3T}\int\frac{d^3p}{(2\pi)^3}\frac{p^2}{\epsilon^2_i} \frac{\omega_{ci}\tau_i}{1+(\omega_{ci}\tau_i)^2}f_{0i},
 \label{sigma_hall}
\end{align}
where $q_{fi}$,  is the fractional electric charge, $\tau_i$ is the thermal averaged relaxation time 
 of the $i$th charged particle species.

\vspace*{0.1cm}

\subsection{Relaxation Time Calculation}
\label{relaxation time}
Relaxation time ($\tau_r$), which is the time required for the system to get back to equilibrium, is defined as the inverse of the relaxation rate \cite{Wiranata:2012br,Plumari:2012ep}. The relaxation rate is the rate of interaction in the system through which momentum of particles gets redistributed towards equilibrium distribution after some small disturbance makes it slightly away from equilibrium.  In presence of strong magnetic field, for the hierarchy $\alpha_s eB<<T^2 \leq eB$, the dominating term for the interaction that is responsible for equilibration of quark and gluons comes from tree level processes such as quark-anti quark annihilation to gluon or vice versa \cite{Hattori:2016lqx}. In this hierarchy, to compare the counter effect of magnetic field and temperature, we consider the maximal effect of magnetic field in the interactions affecting the relaxation process. So we take the relaxation time the same as that in a strong magnetic field.With these assumptions the effective relaxation rate becomes \cite{Kurian:2018dbn,Hattori:2017qih,Fukushima:2017lvb,Cheng:2007jq,Chandra:2012qq,Mitra:2018akk,Kurian:2019fty,Kurian:2018qwb}

\begin{equation}\label{38}
{(\tau_{\text{rel}}^{-1})}=\dfrac{2\alpha_{\text{eff}}C_{F} m_f^2}{\epsilon_{q} 
(1-f_{q}^{0})}\Big (\frac{e^{\mu/T}}{e^{\mu/T}+1}\Big )
( 1+f_{g}^{0})\ln{(T/m_f)}, 
\end{equation}

where $C_F$ is the Casimir factor of the processes and $\alpha_{\text{eff}}$ is the effective coupling constant. ${f}_{q}^{0}=\dfrac{1}{(e^{\beta \epsilon_{i}}+ 1)}$ and 
${f}_{g}^{0}=\dfrac{1}{(e^{\beta \epsilon_{i}}- 1)}$ are respectively the distribution functions for quarks and gluons. In our case, we approximate them to be the Boltzmann distribution function. We take the coupling constant to be explicitly dependent on magnetic field \cite{Rath:2017fdv}

\begin{eqnarray}
\alpha_{eff} &=& \frac{g^2}{4\pi} \nonumber \\
&=&  \frac{1}{{\alpha_s^0(\mu_0)}^{-1}+\frac{11N_c}{12\pi}
\ln\left(\frac{\Lambda_{QCD}^2+M^2_B}{\mu_0^2}\right)+\frac{1}{3\pi}\sum_f \frac{|q_f B|}{\sigma}}
~,\end{eqnarray}

where

\begin{equation}
\alpha_s^0(\mu_0) = \frac{12\pi}
{11N_c\ln\left(\frac{\mu_0^2+M^2_B}{\Lambda_V^2}\right)}
~,\end{equation}
$M_B \sim~1$ GeV is an infrared mass, $\mu_0 = 1.1$ GeV and $\sigma=0.18 ~{\rm{GeV}}^2$ is the string tension.

The above rate is momentum dependent. Taking total relaxation rate of the system with three quark flavors along with their anti-particles, with thermal average of the relaxation rate, one gets the the relevant relaxation, as 
\begin{equation}
\tau=\langle \tau_{rel} \rangle=\frac{1}{ \langle \sum_i \tau_{rel,i}^{-1} \rangle}
\label{tau_av}
\end{equation}

 where 
 \begin{equation}
 \langle \sum_i \tau_{rel,i}^{-1} \rangle=\frac{\sum_i \int d^3p\tau_{rel,i}^{-1} f_{0i}}{\sum_i \int d^3p f_{0i}}
 \label{tau_avfor}
\end{equation}

\section{Results and Discussions}
\label{result}

In this section, we shall discuss our findings regarding electrical conductivity ($\sigma_{el}$) and Hall conductivity ($\sigma_H$) which are obtained by solving the Boltzmann transport equation in presence of magnetic field. We show how the two coefficients vary with temperature and magnetic field for different scenarios. To make this study phenomenologically relevant, we have considered particular collision events. Because the magnetic field depends on centrality of a collision event and the generation of Hall current requires a non-zero baryon chemical potential, we have considered here two particular scenarios for which both the impact parameter dependent magnetic field and the produced baryon chemical potential ($\mu_B$) has been estimated: Pb+Pb collision at $\sqrt {s_{\rm NN}} =2.76$ TeV for which $\mu_B=2$ MeV and Au+Au collision at $\sqrt {s_{\rm NN}} = 200$ GeV for which $\mu_B=25$ MeV ~\cite{Cleymans:2005xv}. Though the field decays either sharply or slowly from the time of formation depending on the conductivity, we take a slightly simplistic assumption for the ease of calculation, that the magnetic field is constant and homogeneous. However, it must be mentioned that this assumption is not very far from the reality when the magnetic field decays slowly. It is worth mentioning that the generation of strong magnetic field happens just after the collisions pertaining to
the pre-equilibrium phase of the spacetime evolution of the system. The quark-gluon plasma being a thermalized matter, the strength of the magnetic field may be lower compared to the initial stages, depending on the conductivity of the system and the equilibration time. 
 We also show the interplay between the $\sigma_{el}$ and $\sigma_H$ for both cases for different magnetic field and temperature. We have assumed the presence of three lightest quark flavors along with their corresponding antiparticles with masses $m_u=3$ MeV, $m_d=5$ MeV and $m_s=100$ MeV.

\begin{figure}[h]
\includegraphics[height=22em]{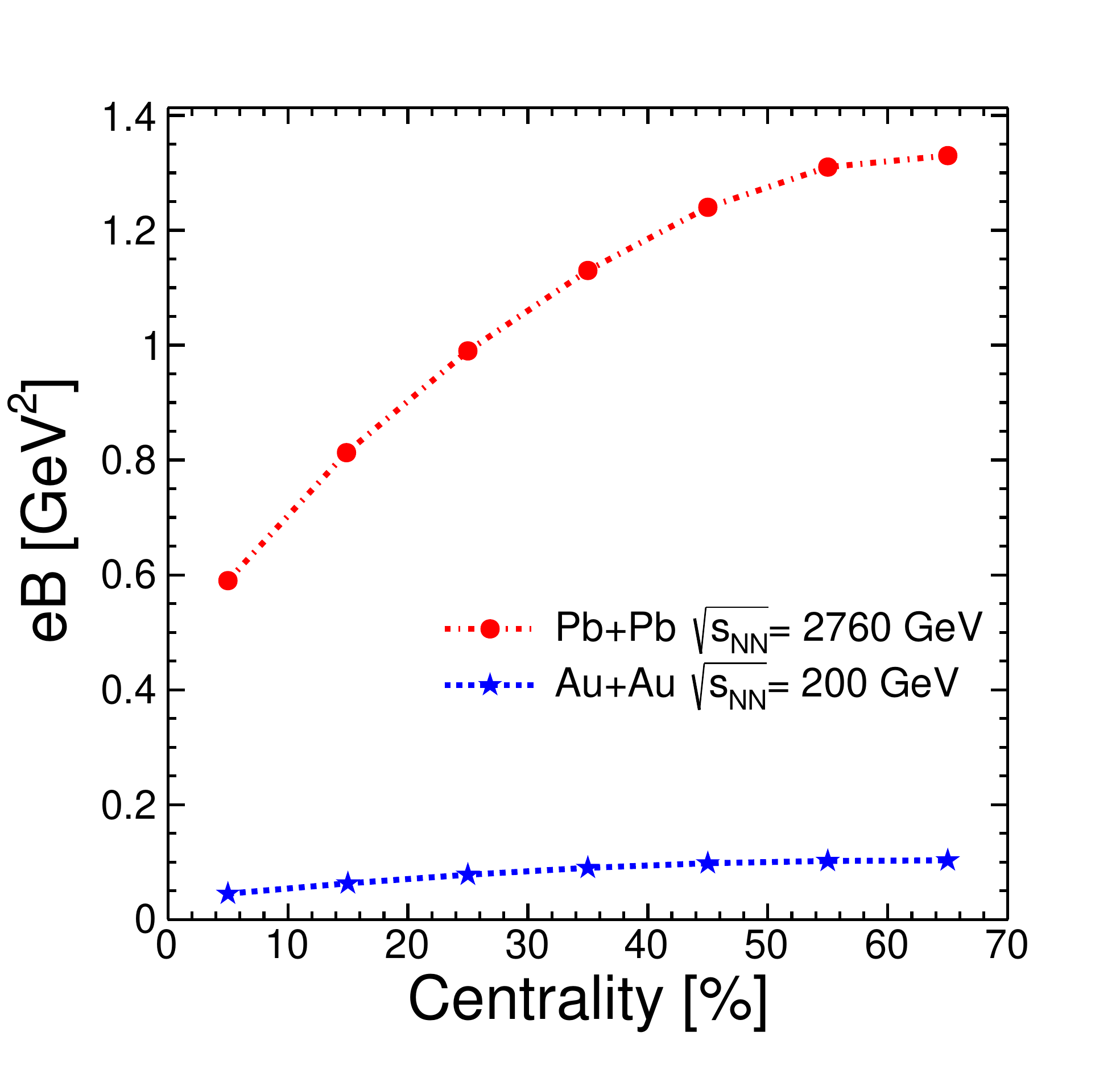}
\caption[]{(Color online) Magnetic field value in different centralities. The values are enlisted in Table. 1}
\label{CentMag}
\end{figure}

In Fig.~\ref{CentMag}, we have shown the centrality dependent magnetic fields for the two experiments we have considered. Magnetic fields for different impact parameters are obtained from ref ~\cite{Hattori:2016emy}. We can see that magnetic field strength increases as the collisions become less and less central. Centrality based impact parameter values are obtained for different collision systems using optical Glauber model ~\cite{Loizides:2017ack}. Then, the corresponding magnetic field values in different centralities has been calculated. After obtaining the value of magnetic fields in different centralities, it is used as an input to calculate the electrical and Hall conductivities for different centrality classes for a specific collision system. We can see in Fig.~\ref{CentMag} that the magnetic field produced in Pb+Pb collision is much higher compared to that produced in Au+Au collision which is expected as higher center-of-mass energy should produce stronger magnetic field.

\begin{figure}[h]
\includegraphics[height=22em]{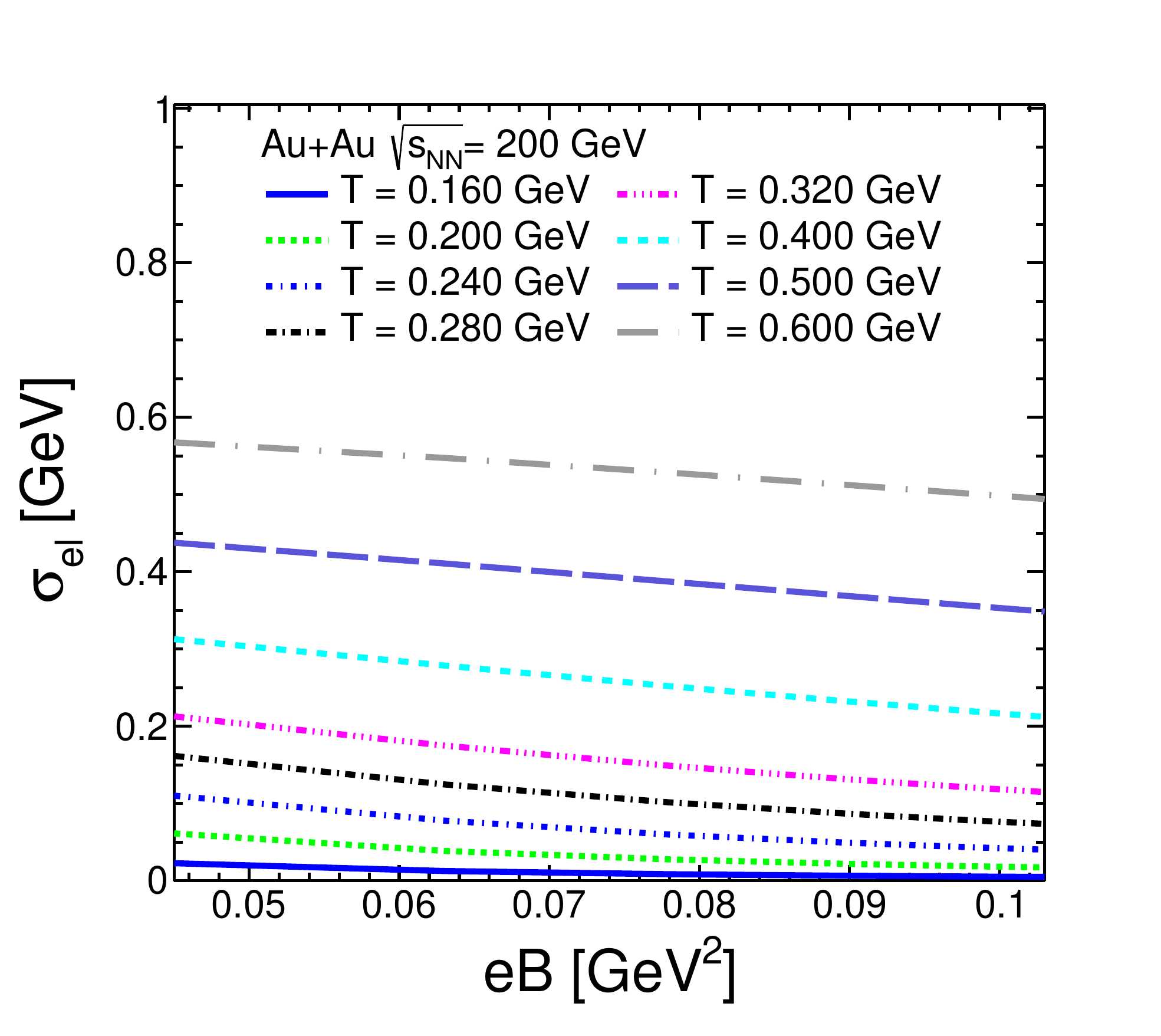}
\caption[]{(Color online) $\sigma_{el}$ vs $eB$ for different temperatures in Au$+$Au collision at $\sqrt {s_{\rm NN}} = 200$ GeV.}
\label{sigmaelAuAuMag}
\end{figure}

\begin{figure}[h]
\includegraphics[height=22em]{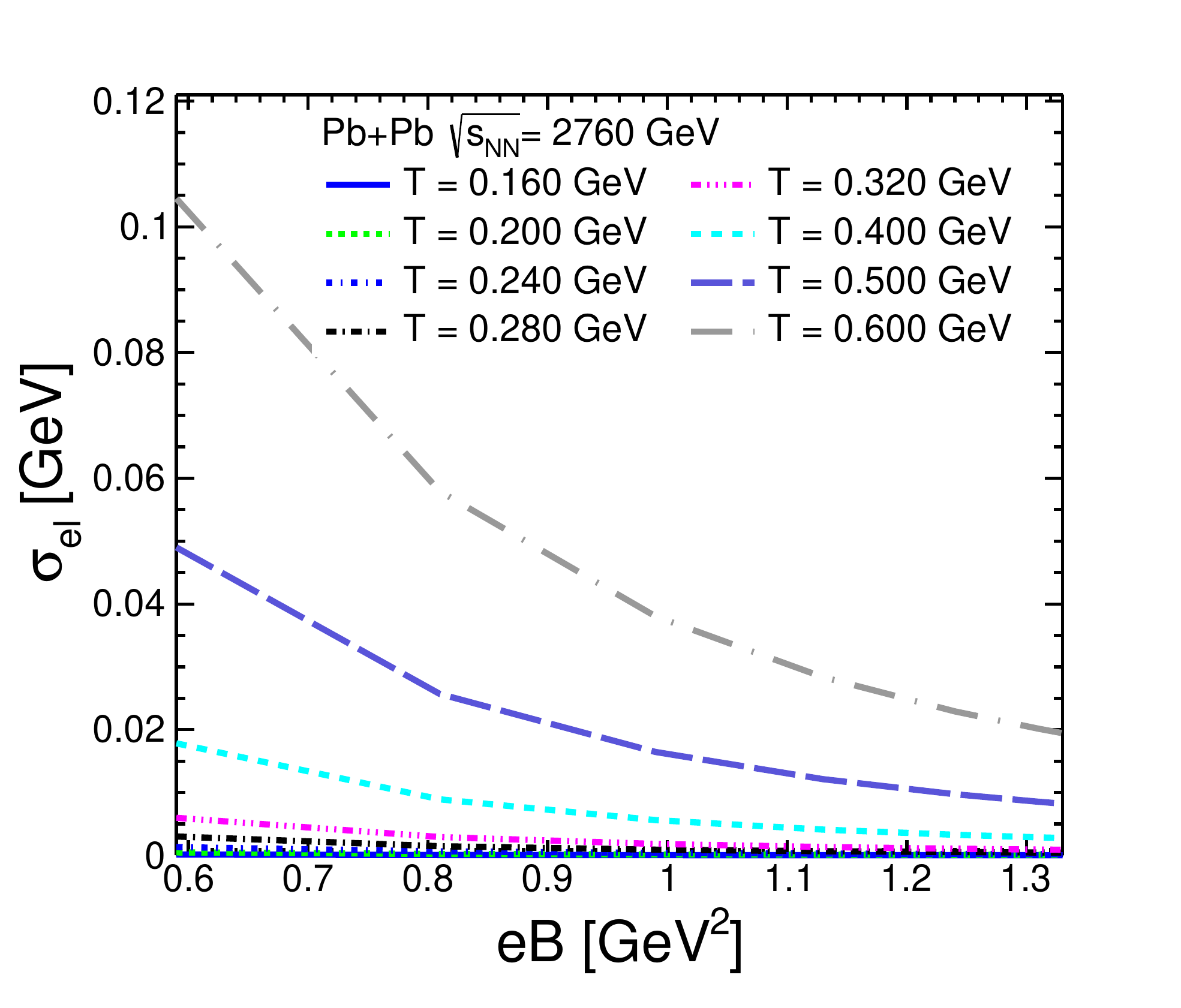}
\caption[]{(Color online) $\sigma_{el}$ vs $eB$ for or different temperatures in Pb$+$Pb collision at $\sqrt {s_{\rm NN}} = 2.76$ TeV.}
\label{sigmaelPbPbMag}
\end{figure}

Fig.~\ref{sigmaelAuAuMag} and Fig.~\ref{sigmaelPbPbMag} shows variation of  $\sigma_{el}$ with $eB$ at different temperatures for both the collision energies we have considered here, i.e. Au+Au and Pb+Pb collisions, respectively.
We can see that $\sigma_{el}$ decreases with increasing magnetic field but increases with temperature for both cases. Also, the fall with magnetic field becomes steeper with increasing temperature. This trend can be understood from eq.\ref{sigma_el}. We can see the relaxation time $\tau_{rel}$ appears in both the numerator and denominator but with a higher power in the denominator. Since $\tau_{rel}$ increases with magnetic field, it is naturally expected for $\sigma_{el}$ to follow a decreasing trend with increasing magnetic field. Also, the cyclotron frequency $\omega_{ci}$, which appears in the denominator, increases with magnetic field and has further contribution in the decline of $\sigma_{el}$. Another reason for this can be that more and more charged particles start moving in a direction perpendicular to the electric field compared to the direction parallel to the field because of increasing Lorentz force as the magnetic field increases which consequently should reduce charge conduction along the electric field. However, this doesn't automatically correspond to an increase in Hall conduction with increasing magnetic field as we shall see later. The increase of $\sigma_{el}$ with temperature is consistent with the findings of previous studies \cite{feng2017,Das:2019wjg} both with and without magnetic field and can also be understood from the eq.\ref{sigma_el}. Since $\tau_{rel}$ decreases with temperature, $\sigma_{el}$ follows the reverse trend.

\begin{figure}[h]
\includegraphics[height=22em]{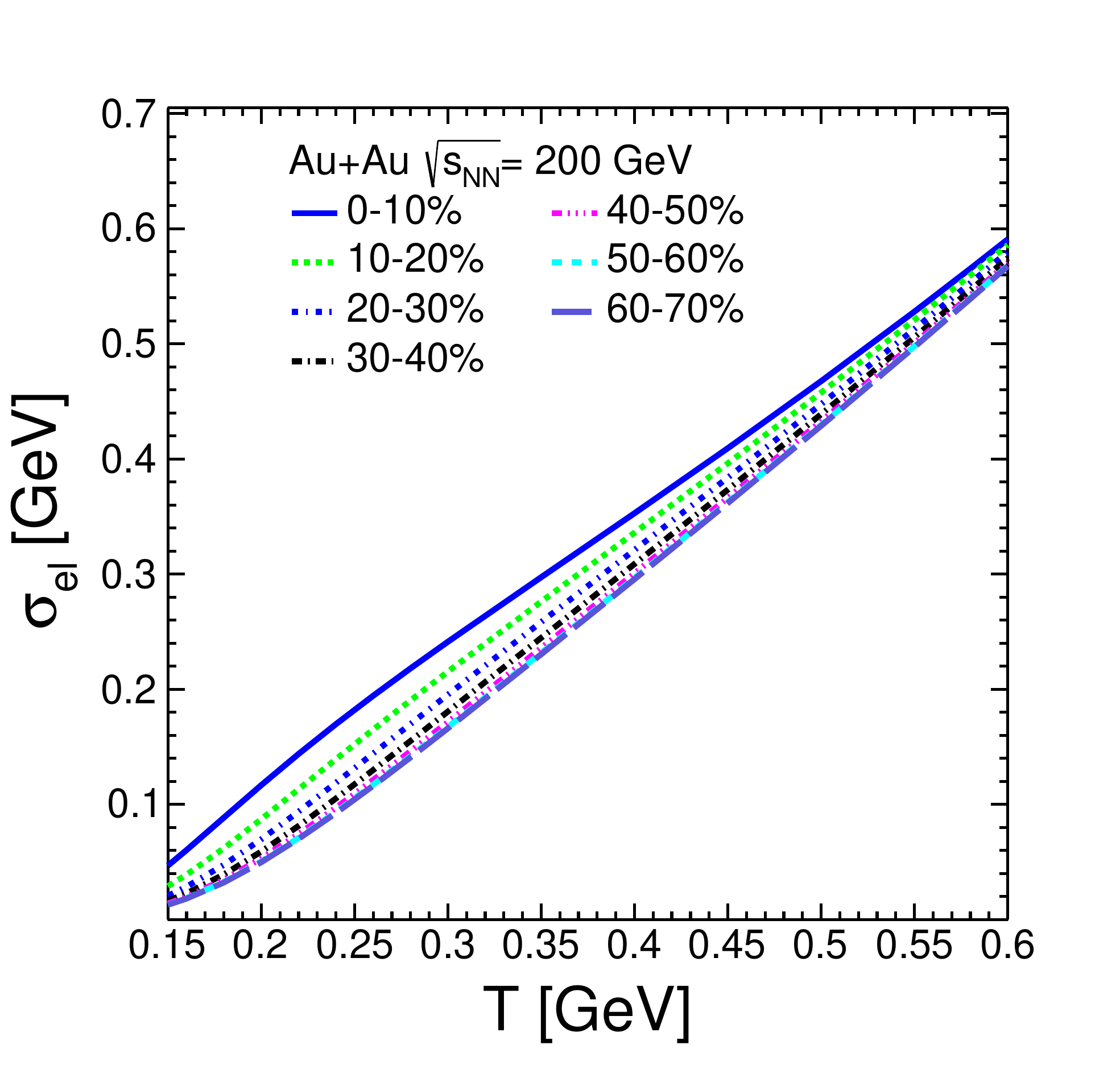}
\caption[]{(Color online) $\sigma_{el}$ vs $T$ for different centrality class in Au$+$Au collision at $\sqrt {s_{\rm NN}} = 200$ GeV.}
\label{sigmaelAuAuTemp}
\end{figure}

\begin{figure}[h]
\includegraphics[height=22em]{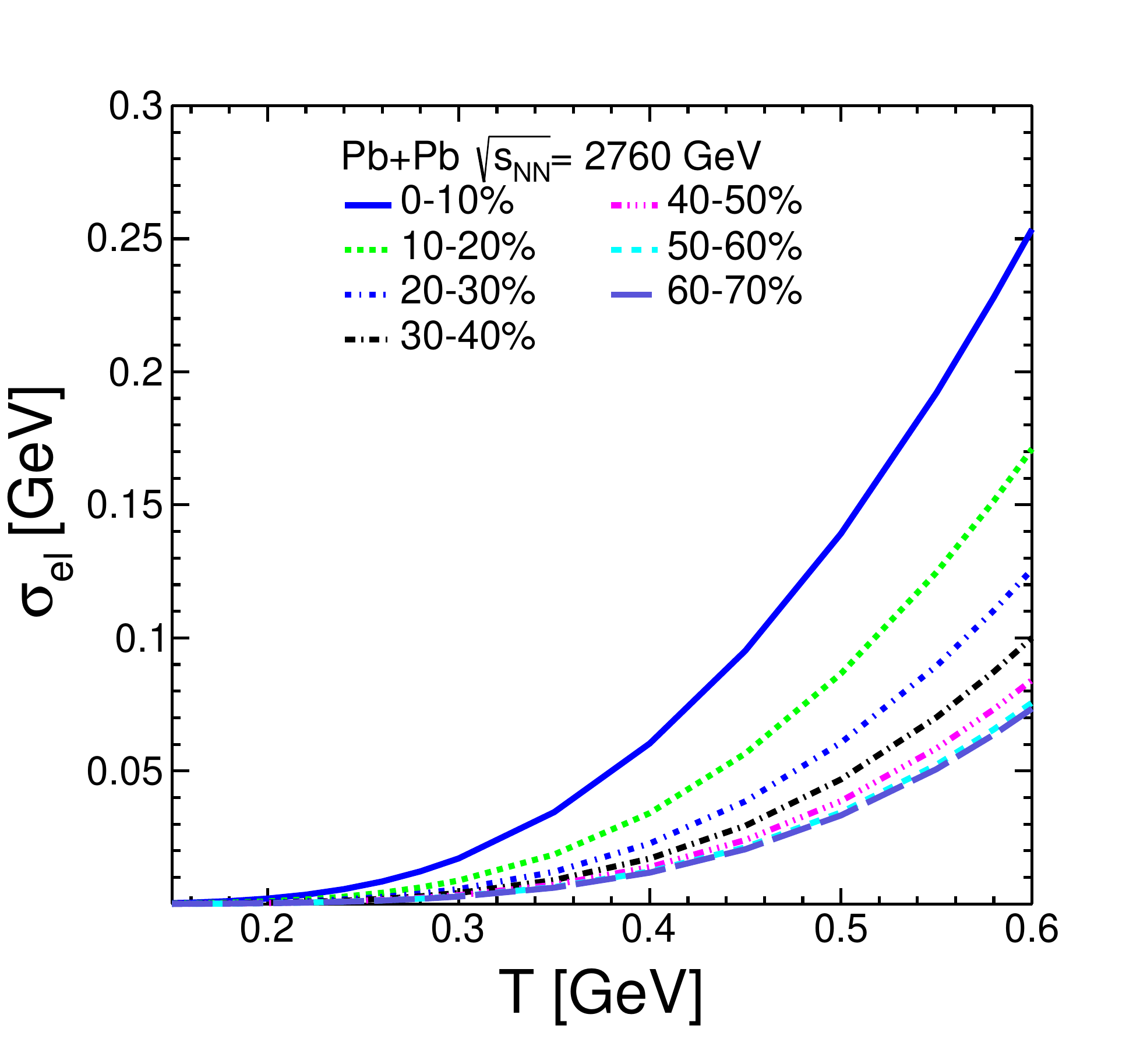}
\caption[]{(Color online) $\sigma_{el}$ vs $T$ for or different centrality class in Pb$+$Pb collision at $\sqrt {s_{\rm NN}} = 2.76$ TeV.}
\label{sigmaelPbPbTemp}
\end{figure}

Fig.~\ref{sigmaelAuAuTemp} and Fig.~\ref{sigmaelPbPbTemp} gives a much clearer picture of the behavior of $\sigma_{el}$ where we have shown the change of $\sigma_{el}$ with temperature for different centrality classes in Au+Au and Pb+Pb collisions, respectively. Since we are considering a static and homogeneous magnetic field, these two figures give us a better idea about what to expect in a particular collision event. As we have mentioned before, different centrality classes correspond to different magnetic field strengths with most central corresponding to lowest value of $eB$ and least central corresponding to highest value of $eB$. In both cases, we see a rapid increase of $\sigma_{el}$ with temperature for all centrality classes with most central (hence lowest magnetic field) case giving the highest values of $\sigma_{el}$. We can see that different centrality affects $\sigma_{el}$ much more prominently in Pb+Pb collision than in Au+Au collision. This makes sense as Pb+Pb collision has a much higher center-of-mass energy ($\sqrt{s}$) compared to Au+Au collision and hence produces a much stronger magnetic field for all centrality classes which in turn creates significant divergence in the values of $\sigma_{el}$ for different centralities. In all the above figures, we can also see that electrical conductivity is much higher in Au+Au collision. This also is because of lower $\sqrt{s}$ and higher $\mu_B$ in Au+Au collision compared to Pb+Pb collision.

So, we can conclude here that electrical conduction is higher in the early phase for both Au+Au and Pb+Pb collisions with magnetic fields lowering the conductivity in both cases and particularly severely for Pb+Pb collision. As the system cools down, electrical conduction will become less for both cases with a steady decline for Au+Au collision and a sharper decline at higher temperature for Pb+Pb collision after which it becomes almost insignificant for all centrality classes near the quark-hadron phase transition. In case of Au+Au collisions, even after the decline, $\sigma_{el}$ will have a significant value near phase transition for all centrality classes. The values of electrical conductivity that we are getting here are significantly different from lattice QCD estimations \cite{Gupta:2003zh} which is to be expected as we are not taking into account the QCD dynamics. But, this result is close to the values obtained in PHSD framework \cite{Cassing:2013iz} .

\begin{figure}[h]
\includegraphics[height=22em]{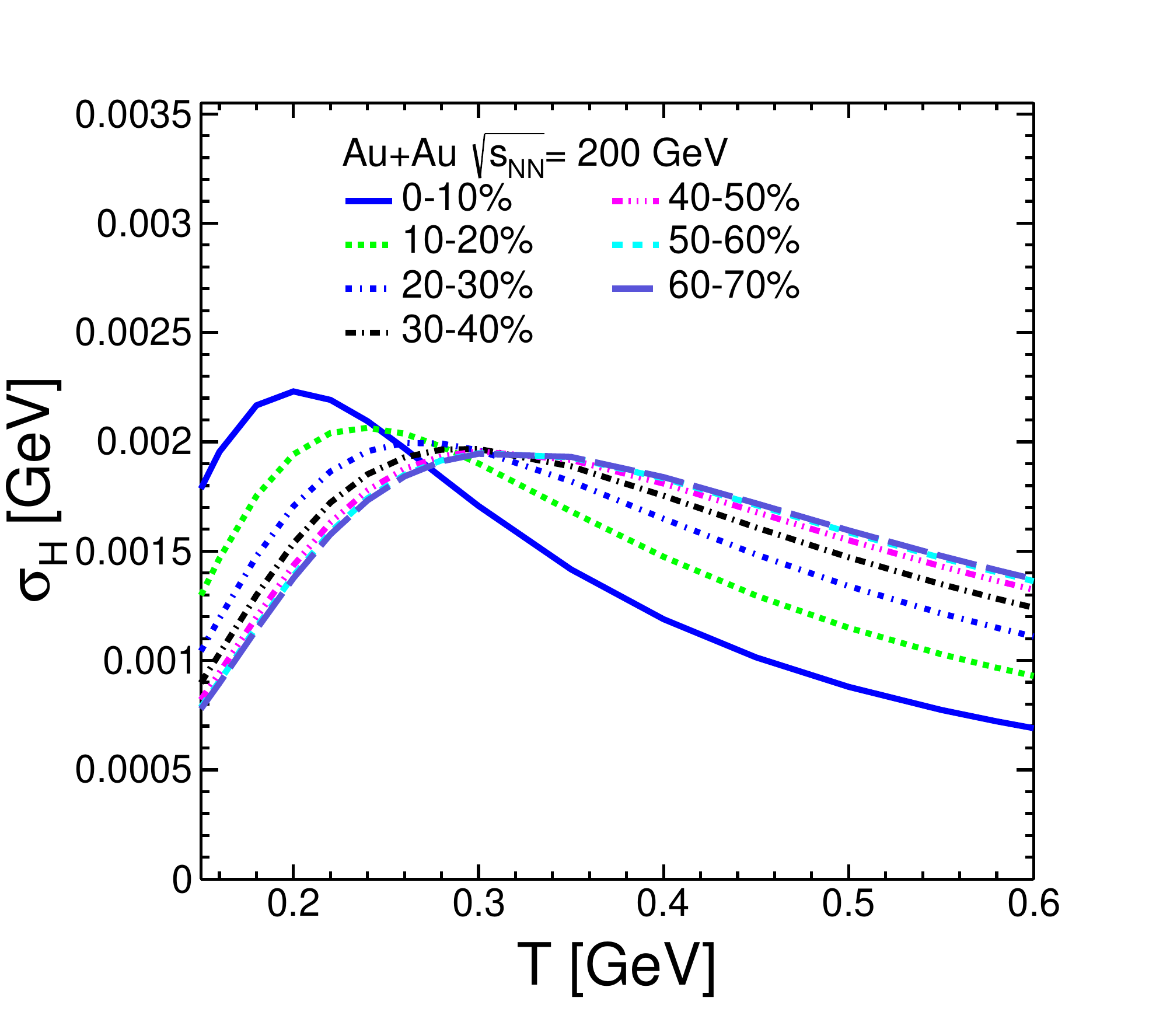}
\caption[]{(Color online) $\sigma_{H}$ vs $T$ for different centrality class in Au$+$Au collision at $\sqrt {s_{\rm NN}} = 200$ GeV.}
\label{sigmahallAuAuTemp}
\end{figure}

\begin{figure}[h]
\includegraphics[height=22em]{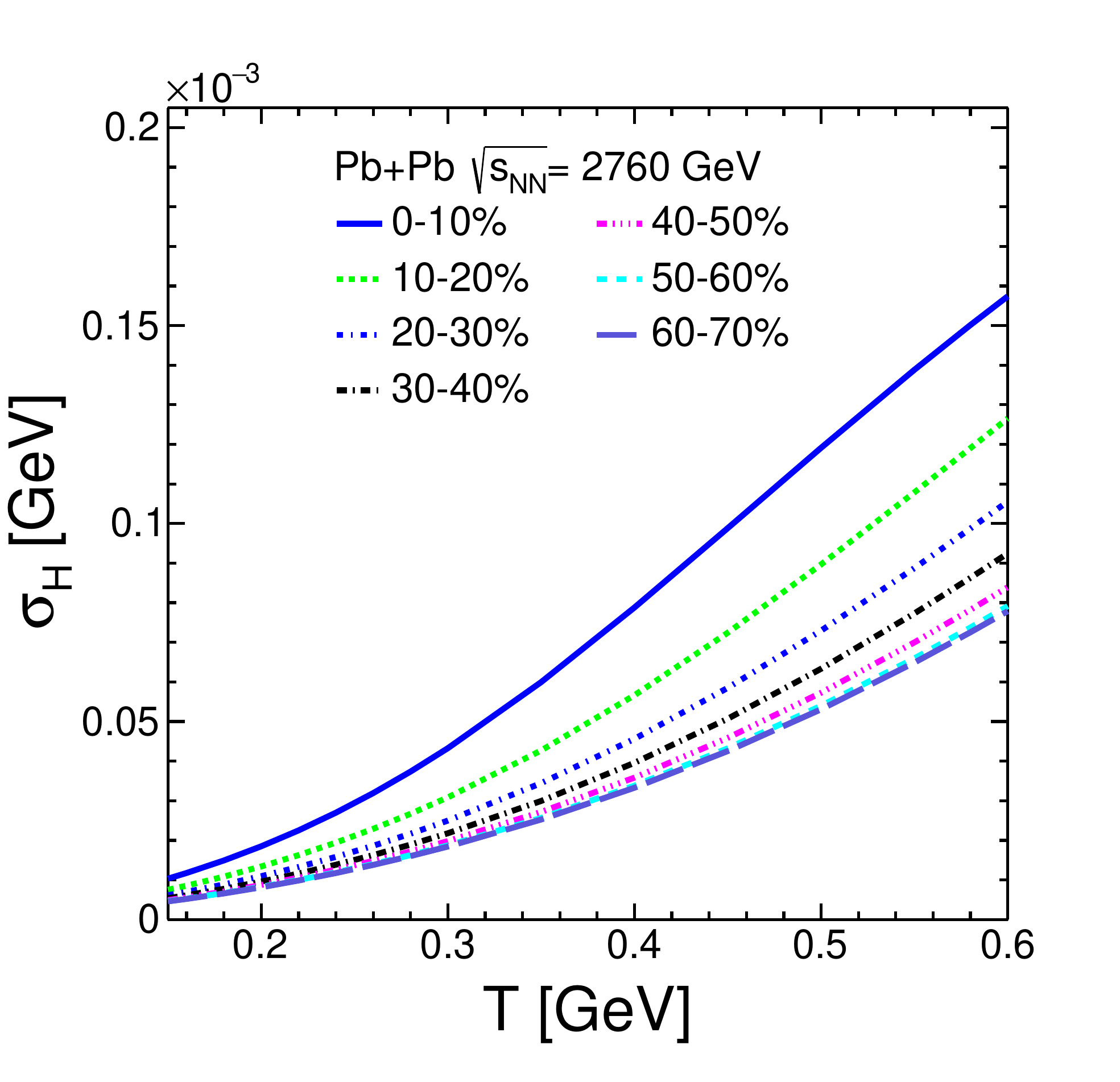}
\caption[]{(Color online) $\sigma_{H}$ vs $T$ for or different centrality class in Pb$+$Pb collision at $\sqrt {s_{\rm NN}} = 2.76$ TeV.}
\label{sigmahallPbPbTemp}
\end{figure}

Next, we show the Hall conductivities for Au+Au and Pb+Pb collisions for different temperatures and magnetic field strengths. Fig.~\ref{sigmahallAuAuTemp} and Fig.~\ref{sigmahallPbPbTemp} shows the variation of $\sigma_H$ with changing temperature for different centrality classes (corresponding to different strengths of magnetic field) respectively for Au+Au and Pb+Pb collisions. First, we notice that the values for Pb+Pb collisions are around an order of magnitude smaller than those for Au+Au collisions. The main reason for this is the very small $\mu_B$ for Pb+Pb collision as the very generation of Hall conduction depends on a finite $\mu_B$ and hence for Pb+Pb collision, Hall current is very small in any situation. We must mention here that the values of $\sigma_H$ are smaller for Au+Au collision also compared to the values of $\sigma_{el}$. We notice that regardless of the temperature or magnetic field strength, $\sigma_H$ is few orders of magnitude lower compared to $\sigma_{el}$. This is because, in realistic scenarios, $\mu_B$ is always small when the magnetic field is very strong at high collision energies. We shall see later that if we consider $\mu_B\sim100$ MeV, the values of $\sigma_H$ becomes comparable to that of $\sigma_{el}$ but as of now, that scenario is not phenomenologically relevant as proposed colliders like FAIR or NICA would likely generate a very small magnetic field (even in the least central events) while producing a large $\mu_B$.

Fig.~\ref{sigmahallAuAuTemp} and Fig.~\ref{sigmahallPbPbTemp} show us that $\sigma_H$ behaves differently with temperature for the two different cases. While for Pb+Pb collision, $\sigma_H$ increases with temperature for all centrality classes, we observe a far more complex behavior for Au+Au collision. We observe reversal of two different trends with increasing temperature for Au+Au collision. First, we see that beyond a certain temperature, $\sigma_H$ starts to decrease for all centralities and then we also observe that while $\sigma_H$ decreases with increasing magnetic field for lower temperature, at higher temperature this trend reverses. To understand this, we need to carefully look at the integrand in eq.\ref{sigma_hall} for two different cases: lower temperature range and higher temperature range. At lower temperature, because of a large relaxation time, the integrand roughly goes as inverse of $\omega_c$ and hence decreases with magnetic field. Also, the temperature behavior is determined by the factor $\frac{f_0}{T}$ which for lower temperature gets dominated by $f_0$ and hence we observe an increasing trend of $\sigma_H$ with temperature. At higher temperature, because of a very small relaxation time, the integrand roughly varies as $\omega_c$ and we observe $\sigma_H$ to be increasing with magnetic field. The temperature behavior in this scenario is determined by the factor $\frac{\tau f_0}{T}$. With a very small value for $\tau$, this term gets dominated by the $\frac{1}{T}$ factor as $f_0$ saturates. So we see a decreasing trend of $\sigma_H$ with temperature. We also observe that the temperature at which this reversal happens increases with magnetic field with a decreasing peak height. The decreasing peak height has also been observed in ~\cite{Das:2019ppb}, though in a different model. In our case, this happens most likely because of our choice of a magnetic field dependent coupling constant which falls with increasing magnetic field.

Fig.~\ref{sigmahallAuAuTemp} also shows that $\sigma_H$ doesn't always increase with magnetic field as one might expect as the Lorentz force increases with magnetic field. Magnetic field generates Hall conduction but at the same time counters it also by putting a directional constraint on the motion of charged particle. In presence of magnetic field, a charged particle goes through a confining circular motion in the plane perpendicular to the field with a radius ($r$) proportional to $\frac{v}{B}$. So, if the velocity perpendicular to the field is low and/or magnetic field strength is very high, the radius of circular motion in the perpendicular plane can be very small. If it becomes smaller than the mean free path ($\lambda$) of the charged particles, then it will adversely affect Hall conduction. So increasing magnetic field can reduce Hall conduction if the velocity of the particle in the perpendicular plane is small. Here, a finite temperature can come to the rescue as it increases the random motion and average velocity of the particle. So, we see that at lower temperature, an increase in temperature increases $\sigma_H$. The peak occurs when $r\sim\lambda$, i.e, the radius of circular motion is roughly equal to the mean free path. Beyond this temperature, $\sigma_H$ decreases as diffusion sets in. Till this region, we see that lower magnetic field produces higher $\sigma_H$ as the radius $r$ is bigger for smaller magnetic field. However, once diffusion sets in, this behavior changes and $\sigma_H$ starts to increase with higher magnetic field. This happens because magnetic field counters diffusion and for very high temperature, very strong magnetic field is required for this.

So, the appearance of peak in Hall conductivity at different temperatures for different strengths of magnetic field can be understood as entirely a result of competition between randomness and increase in average velocity because of temperature and the confining motion induced by magnetic field.

The absence of any change in behavior (and hence any peak) of $\sigma_H$ with temperature in Pb-Pb collisions can also be explained by the above mentioned phenomena. The magnetic field produced in Pb+Pb collisions at $\sqrt {s_{\rm NN}} = $ 2.76 TeV is almost an order of magnitude higher than that of Au+Au at $\sqrt {s_{\rm NN}} =$ 200 GeV. So the temperature range we are considering (0.15 to 0.6 GeV) is not effective at all to counter the confining effect magnetic field, resulting in absence of any peak in this temperature range, as observed for the Au+Au collision.

\begin{figure}[h]
\includegraphics[height=22em]{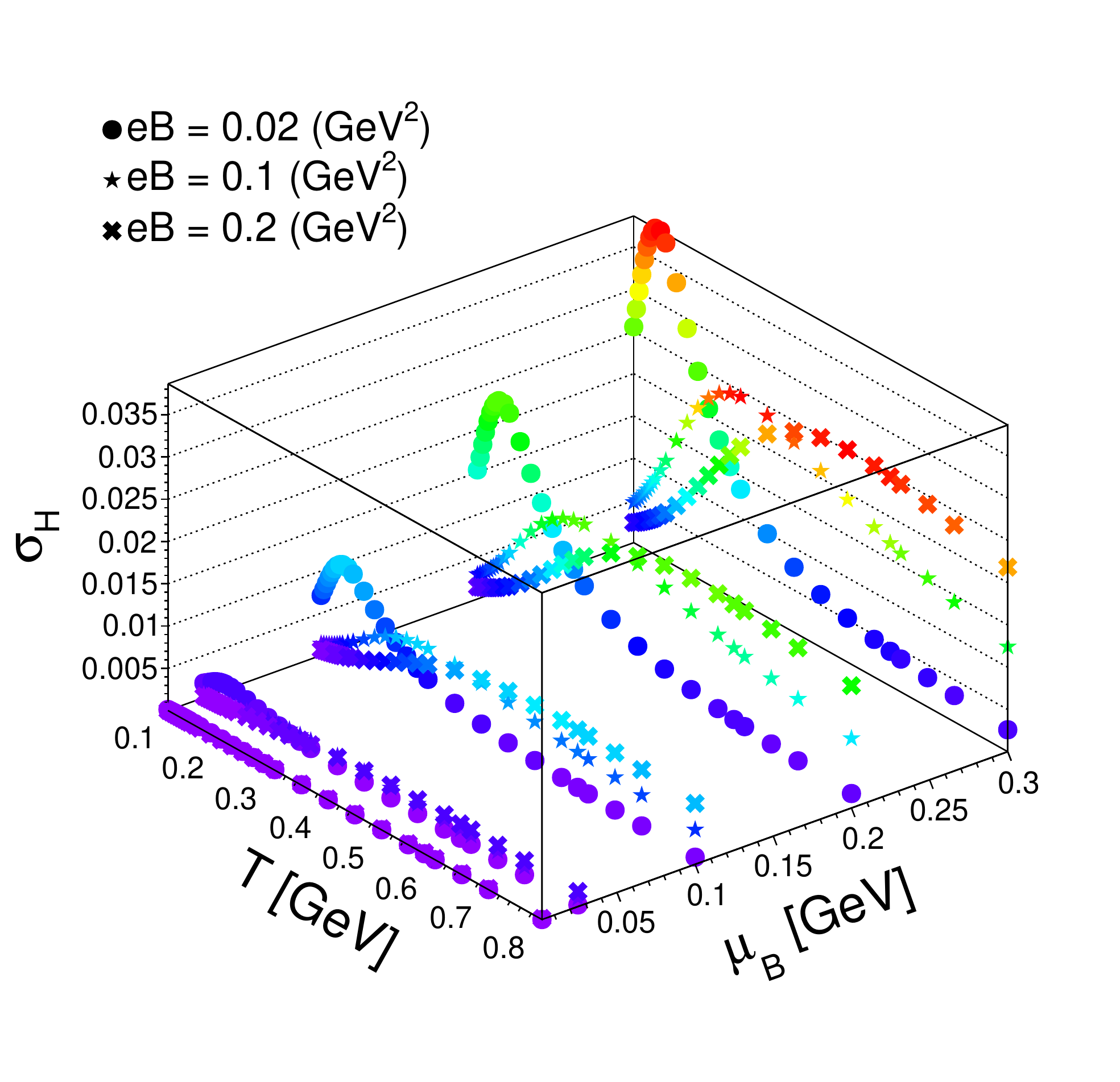}
\caption[]{(Color online) Variation of $\sigma_{H}$ with $T$ and $\mu_B$ for or different magnetic field}
\label{sigmahallTmuB}
\end{figure}

Fig.~\ref{sigmahallTmuB} shows the variation of $\sigma_H$ with both temperature and $\mu_B$ for different magnetic field strengths. In this plot, we have not considered any particular collision scenario but a rather general theoretical situation where a strong magnetic field can coexist with a high value of $\mu_B$. With respect to temperature and magnetic field, we see the similar behavior we observed in Fig.~\ref{sigmahallAuAuTemp} and Fig.~\ref{sigmahallPbPbTemp}. With $\mu_B$, we can observe a clear increase in $\sigma_H$, particularly at lower temperature. At very high $\mu_B$, $\sigma_H$ becomes almost comparable to $\sigma_{el}$.

\begin{figure}[h]
\includegraphics[height=22em]{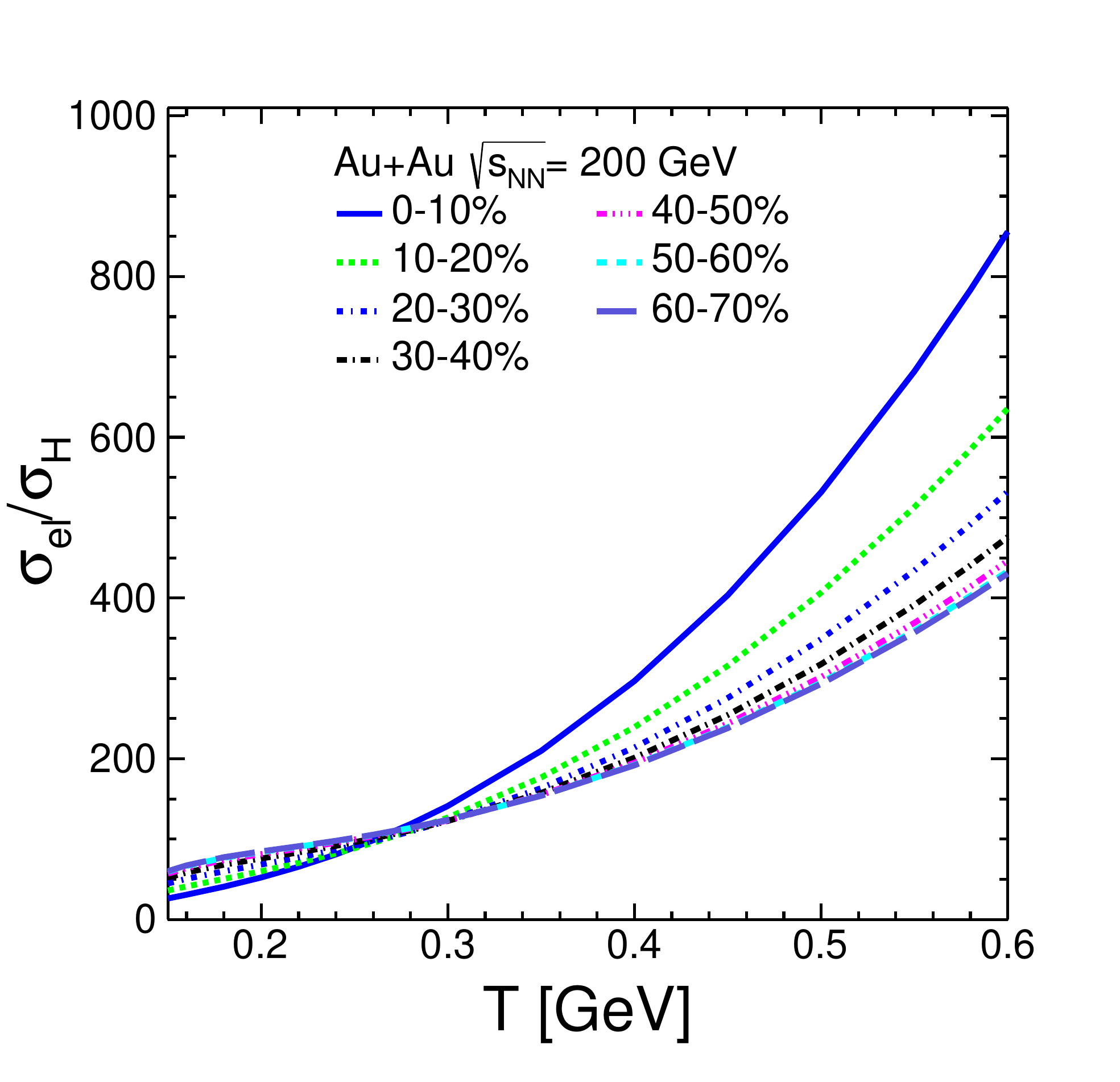}
\caption[]{(Color online) Variation of the ratio $\sigma_{el}/\sigma_{H}$ with $T$ for different centrality classes in Au+Au collision.}
\label{ele_hall_Au}
\end{figure}

In Fig.~\ref{ele_hall_Au}, we show the interplay between the two conductivities for Au+Au collision for different centrality classes. We show how the ratio $\frac{\sigma_{el}}{\sigma_H}$ changes with temperature for different magnetic field strengths. As expected, $\sigma_{el}$ dominates over $\sigma_H$ significantly for most of the region, however, at lower temperature, the dominance is comparatively less and as temperature increases, $\sigma_{el}$ becomes few hundred times larger than $\sigma_H$. This is expected as $\sigma_{el}$ continues to rise throughout the temperature range as shown in Fig.~\ref{sigmaelAuAuTemp} whereas $\sigma_H$ dips at higher temperature. We see here a centrality-wise trend reversal at around $T=250$ MeV. At lower temperature, we see the ratio increases as centrality decreases and this trend reverses at higher temperature. We saw in Fig.~\ref{sigmahallAuAuTemp} that the dip in $\sigma_H$ happens at different temperatures for different centralities, however, for the ratio of the two conductivities, we see that the centrality-wise trend reversal happens at a particular temperature implying a correlation between the two conductivities.

Overall, we can say that in the early stage, when the temperature is very high, electrical conduction will overwhelmingly dominate over Hall conduction making its effects negligible. However, as the medium cools down, the dominance becomes less overwhelming and at lower temperature, closer to the critical point, Hall conduction may have non negligible effects.

\section{Summary and Outlook}
\label{summary}
In this work, we have tried to investigate the electrical and Hall conductivity in hot QGP-like matter in presence of strong magnetic field. We used the Boltzmann transport equation in presence of magnetic field to obtain the expressions for $\sigma_{el}$ and $\sigma_{H}$ using relaxation time approximation. We have incorporated the effects of magnetic field in the relaxation time itself. We have done it by changing the phase space integration in relaxation time as done in ~\cite{Kurian:2018qwb} as well as using a coupling constant which explicitly depends on magnetic field. We have chosen a realistic magnetic field, chemical potential and temperature range to make this study relevant phenomenologically in the context of experiments at RHIC and LHC. We can summarize our findings as follows 

a) The electrical conductivity increases with temperature and decreases with magnetic field. The behavior with temperature is consistent with previous studies and the values we have obtained are also in the same ballpark as previous studies. $\sigma_{el}$ is higher in case of Au+Au collision which is also consistent with previous studies \cite{feng2017,Das:2019wjg}.

b) The Hall conductivity decreases with magnetic field and with temperature it shows a complex behavior. For Pb+Pb collision, it consistently increases with temperature whereas in case of Au+Au collision, it increases at lower temperature and reverses beyond a certain temperature which is dependent on magnetic field. Also, for Au+Au collision, at lower temperature, $\sigma_H$ decreases with magnetic field and reverses the trend at higher temperature. However, the temperature at which the reversal happens is dependent on magnetic field. Also, the value of $\sigma_H$ for a particular centrality is much higher in case of Au+Au collision.

c) $\sigma_{el}$ is always higher compared to $\sigma_H$. At large temperature, Hall conduction is negligible compared to electrical conduction. However, at lower temperature, it can have significant effects. Also, as we have shown in Fig.~\ref{sigmahallTmuB}, at $\mu_B\sim 100$ MeV, the value of $\sigma_H$ can be comparable to that of $\sigma_{el}$.

This work by no means presents a complete picture as we are ignoring the QCD interactions which are the dominant interactions in the QGP phase. So, essentially this study is relevant for a system consisting of nearly massless particles at temperatures and magnetic fields relevant for realistic heavy-ion collisions scenarios. We also have assumed a constant and homogeneous magnetic field which is usually not the case in experiments where the change in magnetic field with time can be rapid as well as moderate depending on the conductivity of the medium. Also, one has to explore the effects of Hall conduction on the observables to create a relatable picture. So, for a more complete picture, the QCD interactions between the particles and the inhomogeneity of the magnetic field should be taken into account. Some of these works are in progress and we will report in future.

\section*{Acknowledgements}
The authors acknowledge the financial supports  from  ALICE  Project  No. SR/MF/PS-01/2014-IITI(G) of Department of Science \& Technology, Government of India. RR acknowledge the financial support by DST-INSPIRE program of Government of India. Authors also thank Dr. Arvind Khuntia for useful discussions on numerical computation.


\begin{thebibliography}{99}


\bibitem{Bazavov:2014pvz} 
  A.~Bazavov {\it et al.} [HotQCD Collaboration],
  Phys.\ Rev.\ D {\bf 90}, 094503 (2014).
  
  \bibitem{Borsanyi:2013bia} 
  S.~Borsanyi, Z.~Fodor, C.~Hoelbling, S.~D.~Katz, S.~Krieg and K.~K.~Szabo,
  Phys.\ Lett.\ B {\bf 730}, 99 (2014).

\bibitem{Krzewicki:2011ee} 
  M.~Krzewicki [ALICE Collaboration],
  J.\ Phys.\ G {\bf 38}, 124047 (2011).

\bibitem{Aamodt:2010cz} 
  K.~Aamodt {\it et al.} [ALICE Collaboration],
  Phys.\ Rev.\ Lett.\  {\bf 106}, 032301 (2011).
  
  \bibitem{Hirano:2010je} 
  T.~Hirano, P.~Huovinen and Y.~Nara,
  Phys.\ Rev.\ C {\bf 84}, 011901 (2011).
  
  \bibitem{Arsene:2004fa} 
  I.~Arsene {\it et al.} [BRAHMS Collaboration],
  Nucl.\ Phys.\ A {\bf 757}, 1 (2005).
  
  \bibitem{Back:2004je} 
  B.~B.~Back {\it et al.},
  Nucl.\ Phys.\ A {\bf 757}, 28 (2005).
  
  \bibitem{Adams:2005dq} 
  J.~Adams {\it et al.} [STAR Collaboration],
  Nucl.\ Phys.\ A {\bf 757}, 102 (2005).
  
  \bibitem{Shuryak:2004cy} 
  E.~V.~Shuryak,
  Nucl.\ Phys.\ A {\bf 750}, 64 (2005).
  \bibitem{Gyulassy:2004zy} 
  M.~Gyulassy and L.~McLerran,
  Nucl.\ Phys.\ A {\bf 750}, 30 (2005).
  
  \bibitem{Muller:2006ee} 
  B.~Muller and J.~L.~Nagle,
  Ann.\ Rev.\ Nucl.\ Part.\ Sci.\  {\bf 56}, 93 (2006).
  
\bibitem{Kovtun:2004de} 
  P.~Kovtun, D.~T.~Son and A.~O.~Starinets,
  Phys.\ Rev.\ Lett.\  {\bf 94}, 111601 (2005).

\bibitem{Romatschke:2007mq} 
  P.~Romatschke and U.~Romatschke,
  Phys.\ Rev.\ Lett.\  {\bf 99}, 172301 (2007).

\bibitem{Heinz:2013th} 
  U.~Heinz and R.~Snellings,
  Ann.\ Rev.\ Nucl.\ Part.\ Sci.\  {\bf 63}, 123 (2013).
  
  \bibitem{Gale:2013da} 
  C.~Gale, S.~Jeon and B.~Schenke,
  Int.\ J.\ Mod.\ Phys.\ A {\bf 28}, 1340011 (2013).
  
  \bibitem{Teaney:2009qa} 
  D.~A.~Teaney, arXiv:0905.2433 [nucl-th]. 
  
  \bibitem{Romatschke:2009im} 
  P.~Romatschke,
  Int.\ J.\ Mod.\ Phys.\ E {\bf 19}, 1 (2010).

\bibitem{Luzum:2008cw} 
  M.~Luzum and P.~Romatschke,
  Phys.\ Rev.\ C {\bf 78}, 034915 (2008)
  Erratum: [Phys.\ Rev.\ C {\bf 79}, 039903 (2009)].
  
\bibitem{Song:2007ux} 
  H.~Song and U.~W.~Heinz,
  Phys.\ Rev.\ C {\bf 77}, 064901 (2008).
  
  \bibitem{Dusling:2007gi} 
  K.~Dusling and D.~Teaney,
  Phys.\ Rev.\ C {\bf 77}, 034905 (2008).
  
  \bibitem{Molnar:2008xj} 
  D.~Molnar and P.~Huovinen,
  J.\ Phys.\ G {\bf 35}, 104125 (2008).
  
  \bibitem{Bozek:2009dw} 
  P.~Bozek,
  Phys.\ Rev.\ C {\bf 81}, 034909 (2010).

\bibitem{Chaudhuri:2009hj} 
  A.~K.~Chaudhuri,
  J.\ Phys.\ G {\bf 37}, 075011 (2010).
  
  \bibitem{Schenke:2010rr} 
  B.~Schenke, S.~Jeon and C.~Gale,
  Phys.\ Rev.\ Lett.\  {\bf 106}, 042301 (2011).
  
  \bibitem{Skokov:2009qp} 
  V.~Skokov, A.~Y.~Illarionov and V.~Toneev,
  Int.\ J.\ Mod.\ Phys.\ A {\bf 24}, 5925 (2009).
  
\bibitem{Hattori:2016emy} 
  K.~Hattori and X.~G.~Huang,
  Nucl.\ Sci.\ Tech.\  {\bf 28}, 26 (2017).

 \bibitem{Kharzeev:2007jp} 
  D.~E.~Kharzeev, L.~D.~McLerran and H.~J.~Warringa,
  Nucl.\ Phys.\ A {\bf 803}, 227 (2008).
  
   \bibitem{Tuchin:2013ie} 
  K.~Tuchin,
  Adv.\ High Energy Phys.\  {\bf 2013}, 490495 (2013).
  
\bibitem{Kharzeev:2012ph} 
  D.~E.~Kharzeev, K.~Landsteiner, A.~Schmitt and H.~U.~Yee,
  Lect.\ Notes Phys.\  {\bf 871}, 1 (2013).

\bibitem{Ding:2010ga}
H.~T.~Ding, A.~Francis, O.~Kaczmarek, F.~Karsch, E.~Laermann and W.~Soeldner,
Phys. Rev. D \textbf{83}, 034504 (2011).

\bibitem{Moore:2006qn}
G.~D.~Moore and J.~M.~Robert,
arXiv:hep-ph/0607172 [hep-ph].


\bibitem{feng2017}
B. Feng, Phys. Rev. {\bf D96}, 036009 (2017).

 \bibitem{Das:2019wjg} 
  A.~Das, H.~Mishra and R.~K.~Mohapatra,
  Phys.\ Rev.\ D {\bf 99}, 094031 (2019).
  
  \bibitem{Das:2019ppb} 
  A.~Das, H.~Mishra and R.~K.~Mohapatra,
  Phys.\ Rev.\ D {\bf 101}, 034027 (2020).
    
\bibitem{Hosoya:1983xm} 
  A.~Hosoya and K.~Kajantie,
  Nucl.\ Phys.\ B {\bf 250}, 666 (1985).
    
  \bibitem{Wiranata:2012br} 
  A.~Wiranata and M.~Prakash,
  Phys.\ Rev.\ C {\bf 85}, 054908 (2012).
 
\bibitem{Plumari:2012ep} 
  S.~Plumari, A.~Puglisi, F.~Scardina and V.~Greco,
  Phys.\ Rev.\ C {\bf 86}, 054902 (2012).

\bibitem{Hattori:2016lqx} 
  K.~Hattori, S.~Li, D.~Satow and H.~U.~Yee,
  Phys.\ Rev.\ D {\bf 95}, 076008 (2017).
  
  \bibitem{Kurian:2018dbn} 
  M.~Kurian and V.~Chandra,
  Phys.\ Rev.\ D {\bf 97}, 116008 (2018).
  
  
   \bibitem{Hattori:2017qih} 
  K.~Hattori, X.~G.~Huang, D.~H.~Rischke and D.~Satow,
  Phys.\ Rev.\ D {\bf 96}, 094009 (2017).
  \bibitem{Fukushima:2017lvb} 
  K.~Fukushima and Y.~Hidaka,
  Phys.\ Rev.\ Lett.\  {\bf 120}, 162301 (2018).
  \bibitem{Cheng:2007jq} 
  M.~Cheng {\it et al.},
  Phys.\ Rev.\ D {\bf 77}, 014511 (2008).   
  \bibitem{Chandra:2012qq} 
  V.~Chandra,
  Phys.\ Rev.\ D {\bf 86}, 114008 (2012).
  \bibitem{Mitra:2018akk} 
  S.~Mitra and V.~Chandra,
  Phys.\ Rev.\ D {\bf 97}, 034032 (2018).    
  \bibitem{Kurian:2019fty} 
  M.~Kurian and V.~Chandra,
  Phys.\ Rev.\ D {\bf 99}, 116018 (2019).  
  
  
  \bibitem{Kurian:2018qwb} 
  M.~Kurian, S.~Mitra, S.~Ghosh and V.~Chandra,
  Eur.\ Phys.\ J.\ C {\bf 79}, 134 (2019).
  
  \bibitem{Rath:2017fdv} 
  S.~Rath and B.~K.~Patra,
  JHEP {\bf 1712}, 098 (2017).
  
 \bibitem{Cleymans:2005xv} 
  J.~Cleymans, H.~Oeschler, K.~Redlich and S.~Wheaton,
  Phys.\ Rev.\ C {\bf 73}, 034905 (2006).
  

  \bibitem{Loizides:2017ack} 
  C.~Loizides, J.~Kamin and D.~d'Enterria,
  Phys.\ Rev.\ C {\bf 97}, 054910 (2018)
  Erratum: [Phys.\ Rev.\ C {\bf 99}, 019901 (2019)].
  
\bibitem{Cassing:2013iz} 
W.~Cassing, O.~Linnyk, T.~Steinert and V.~Ozvenchuk, Phys.\ Rev.\ Lett. {\bf 110}, 182301 (2013).


\bibitem{Gupta:2003zh} 
S.~Gupta, Phys.\ Lett.\ B {\bf 597}, 57 (2004).

  
 

  
 
    
  
  
  
  
     
  
\end{thebibliography}
\end{document}